\documentclass{iaus}
\usepackage{graphics}

  \checkfont{eurm10}
  \iffontfound
    \IfFileExists{upmath.sty}
      {\typeout{^^JFound AMS Euler Roman fonts on the system,
                   using the 'upmath' package.^^J}%
       \usepackage{upmath}}
      {\typeout{^^JFound AMS Euler Roman fonts on the system, but you
                   dont seem to have the}%
       \typeout{'upmath' package installed. iaus.cls can take advantage
                 of these fonts,^^Jif you use 'upmath' package.^^J}%
      }
  \else
  \fi

  \checkfont{msam10}
  \iffontfound
    \IfFileExists{amssymb.sty}
      {\typeout{^^JFound AMS Symbol fonts on the system, using the
                'amssymb' package.^^J}%
       \usepackage{amssymb}%

      }{}
  \fi

  \IfFileExists{amsbsy.sty}
    {\typeout{^^JFound the 'amsbsy' package on the system, using it.^^J}%
     \usepackage{amsbsy}}
    {\providecommand\boldsymbol[1]{\mbox{\boldmath $##1$}}}

\newsavebox{\astrutbox}
\sbox{\astrutbox}{\rule[-5pt]{0pt}{20pt}}

\title[Outskirts of Galaxy Clusters: intense life in the suburbs]
      {The roles of ram-pressure stripping and minor mergers in evolution of galaxies}

\author[T. Okamoto \and M. Nagashima]%
{Takashi Okamoto$^1$ $^2$%
\and Masahiro Nagashima$^3$ $^1$}

\affiliation{$^1$Department of Physics, University of Durham, South Road, Durham, DH1 3LE UK: takashi.okamoto@durham.ac.uk\\[\affilskip]
$^2$NAOJ, NINIS, Mitaka, Tokyo 181-8588 Japan\\[\affilskip]
$^3$Department of Physics, Kyoto University, Kyoto 606-8502, Japan}

\pubyear{2004}
\volume{195}
\pagerange{1--5}
\date{?? and in revised form ??}
\jname{Outskirts of Galaxy Clusters: intense life in the suburbs}
\editors{A. Diaferio, ed.}
\begin{document}

\maketitle

\begin{abstract}
We investigate environmental effects on evolution of bright cluster
galaxies in a $\Lambda$-dominated cold dark matter universe
using a combination of dissipationless $N$-body simulations and a
semi-analytic galaxy formation model.  
We incorporate effects of ram-pressure stripping (RPS) and minor 
merger-induced small starburst (minor burst) into our model. 
By considering minor burst, observed morphology-radius relation is 
successfully reproduced. When we do not consider minor burst, 
the RPS hardly increases the intermediate B/T population. 
In addition, the RPS and minor burst are not important for colours or 
star formation rates of galaxies in the cluster core if star formation 
time-scale is properly chosen, because the star formation  is sufficiently 
suppressed by consumption of the cold gas.  
We also find that SF in bulge-dominated galaxies is mainly 
terminated by starburst induced by major mergers in all environments.
\end{abstract}

\firstsection 
\section{Introduction}

It has been found that galaxy morphology is a function of environment
(Dressler 1980; Whitmore et al. 1993) and redshift (Dressler et al. 1997). 
Colours and star formation rate (SFRs) of galaxies also show similar 
dependence on environment and redshift (e.g. Butcher \& Oemler 1984; 
Lewis et al. 2002).  
To account for these observational trends, several mechanisms that may suppress 
the star formation (SF) and transform one morphological type into another have been 
proposed.

Interaction between galaxies is one possible process to promote morphological 
transformation. Numerical simulations confirmed that major mergers produce 
galaxies resembling ellipticals as merger remnants (Barnes 1996) and that 
accretion of small satellites onto their host spiral lead a host spiral to 
S0 type (Walker et al. 1996). 
Because the galaxy merger triggers starburst, the cold gas contained in 
original galaxies is exhausted in a very short time. 

The second mechanism is removal of hot reservoir. 
In denser environment such as clusters, diffuse hot gas reservoir that is originally 
confined in haloes of non-central galaxies becomes part of intracluster medium. 
A galaxy whose hot gas reservoir is removed slowly exhausts its cold gas in 
its SF time-scale (strangulation). 

Above 2 processes, i.e. major merger-induced starburst and strangulation, 
has been incorporated in most of semi-analytic (SA) models which have failed to 
reproduce intermediate $B/T$ population (Okamoto \& Nagashima 2001; Diaferio et al. 2001). 
In this paper, we introduce two additional processes in order to solve this problem.
One is ram-pressure stripping (RPS) of the cold gas from galactic disks 
(Gunn \& Gott 1972).
Another is minor merger-induced small 
starburst. Hydrodynamic simulations showed some fraction of disk gas is 
fuelled to the galactic centre by a minor merger (Walker et al. 1996). 
By using a combination of $N$-body simulations with a SA model, 
we study roles of above 4 processes in evolution of galaxies and what process 
is responsible for 
the formation of intermediate $B/T$ galaxies. 

\section{Models}

We here consider a $\Lambda$-dominated cold dark matter universe
($\Omega_0=0.3, \ \lambda_0=0.7, \ \Omega_b=0.015 h^{-2}, \ 
h = 100 {\rm km s}^{-1} {\rm Mpc}^{-1} = 0.7$, and $\sigma_8 = 1$). 
We simulate a cluster region to study cluster galaxies and 
a region whose density is almost the same as the mean density of the 
universe to study field population.  We then construct merger trees of 
virialised haloes and their substructure haloes. The details of this 
procedure is given in Okamoto \& Habe (1999, 2000). 

To clarify the effect of each physical process, 
we here examine 4 models. The model in which we do not consider 
either the RPS or the minor burst is called the {\it standard} 
model, because this is the standard prescription in SA models. 
Model parameters in the standard model is normalised so that 
it reproduces luminosity functions, cold gas fractions, and sizes 
of field galaxies. 
We also show the model using 4 times longer SF time-scale 
than the standard model to see the effect of the strangulation. 
This model is referred to as the {\it low SFR} model. 
The model with the minor burst and the model with the RPS are 
called "minor burst" and "RPS", respectively. 
In these models, we use the same parameters as used in 
the standard model. The details can be found in Okamoto \& Nagashima (2003). 

\section{Results}
\begin{figure}[thb]
\begin{center}
\scalebox{0.4}{
\includegraphics{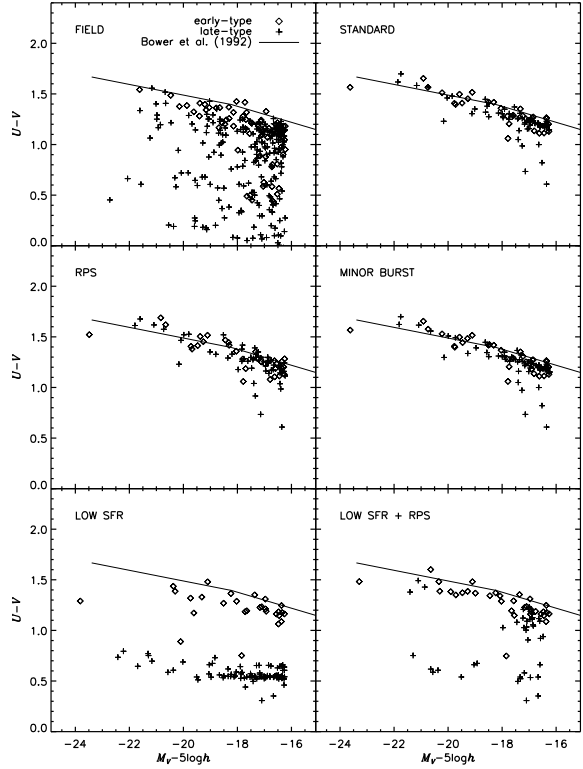}
}
\end{center}
\caption{CMRs at $z = 0$. 
We show $U-V$ colours of galaxies in the field model and galaxies within the cluster core in the standard , RPS, minor burst, low SFR, and low SFR + RPS models as a function of $V$-band luminosity. 
Early-type galaxies and late-type galaxies are represented by diamonds and pluses, respectively. The solid lines show the observed CMR for cluster ellipticals by Bower et al. (1992).
\label{fig1}}
\end{figure}

\begin{figure}[thb]
\begin{center}
\scalebox{0.23}{
\includegraphics{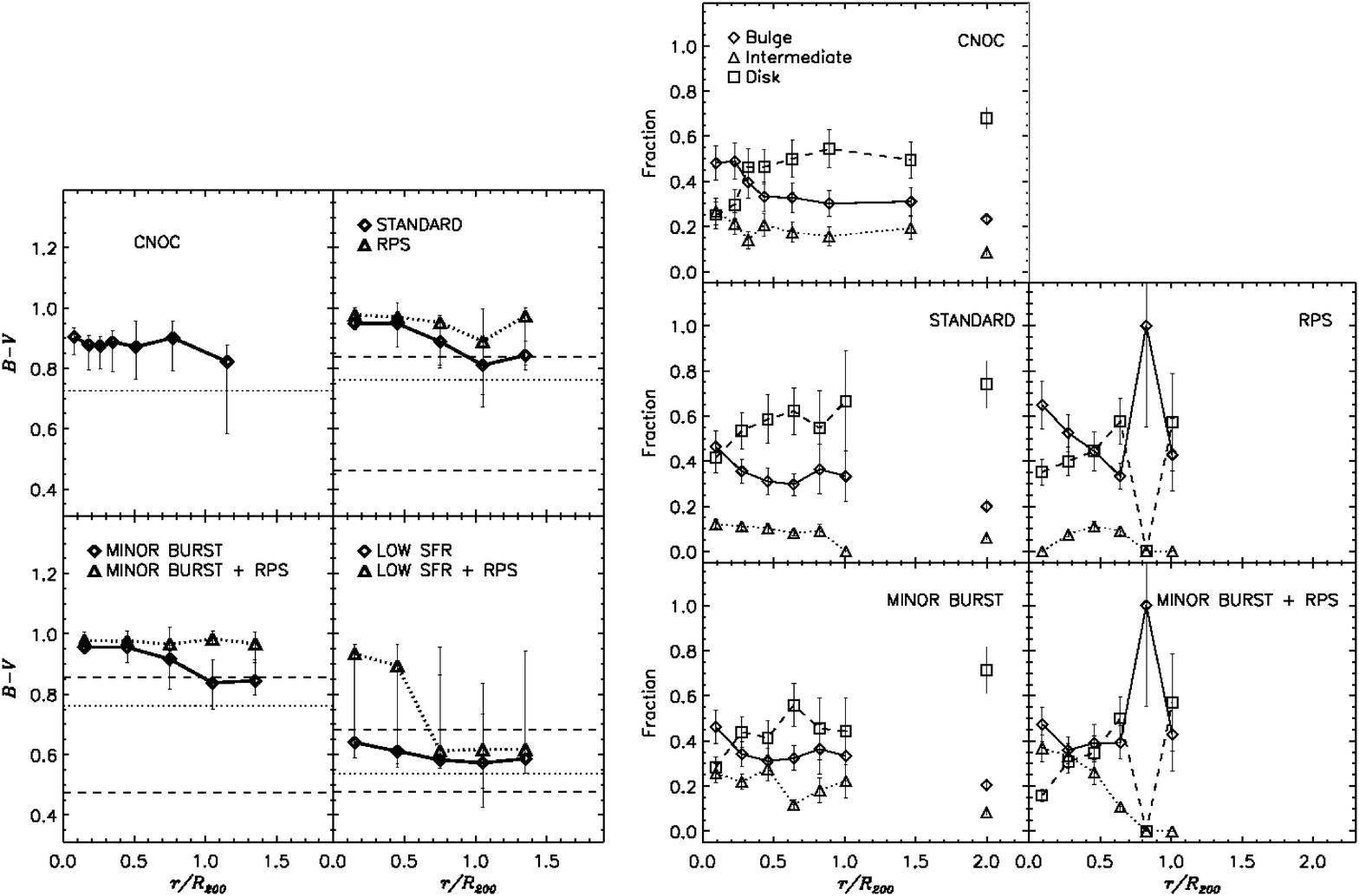}
}
\end{center}
\caption{
Left: The median $B$-$V$ colours of cluster galaxies with $M_R < -20.5$ are plotted as a function of projected radius. In the upper left panel, the diamonds denote the observed median colours of the CNOC sample.
The median calor of field galaxies at the same redshift range are shown by the dotted line.
Other panels shows the median colours for the simulated galaxies at $z = 0.2$ and the corresponding RPS model is indicated by the triangles in each panel. The field model is also shown by the dotted line in each panel. 
Right: The morphological fractions are plotted as a function of projected radius.
In the top panel we show the observed fractions for the galaxies with $M_R < -20.5$ in the CNOC sample.  
Other panels show the fractions given by our models for the simulated cluster at $z = 0.2$.  
The symbols at $r/R_{200} = 2$ indicate the fractions in the field in each panel. 
\label{fig2}}
\end{figure}
In Fig. \ref{fig1}, we show colour-magnitude relations (CMRs) for simulated galaxies in the standard 
field model and cluster cores ($ r < 0.5 h^{-1}$ Mpc). We classify galaxies with $B$-band $B/T < 0.4$ 
as late-type and others as early-type galaxies. First of all, the CMRs for early-type galaxies are quite 
similar in all models, even in the field. It suggests that major merger-induced starbursts mainly truncate
SF in early-type galaxies. On the other hand, colours of late-type galaxies are sensitive to 
the SF time-scale and their environments. Blue population seen in the field model is vanished 
in the cluster if we use appropriate SF time-scale. Since, in the low SFR model, late-type galaxies 
have much bluer colours than those in the standard model, it can be said that strangulation is the process 
that stops  SF of late-type galaxies in cluster cores. The suppression of SF by strangulation is so efficient 
that the effect of the RPS is hardly seen except in the low SFR + RPS model.

It has been known that galaxies in inner regions of clusters are redder than those in outer regions. 
Recent observations have confirmed the existence of colour gradients in the galaxy population, which 
appear to continue even beyond the virial radius (e.g. Carlberg et al. 1997). 
In the left panels of Fig. \ref{fig2} we show the $B-V$ colours of the cluster galaxies at $z = 0.2$ as a function 
of projected radius. The projected radius is normalised by $R_{200}$ that is the radius inside which 
the mean density is 200 times the critical density of the universe at a given redshift. 
The observational data are taken from Diaferio et al. (2001). 

For the standard and minor burst models, the effect of the RPS can be seen only at large radii, 
because, in the central part, the SF is strongly suppressed by the strangulation. 
By comparing the low SFR + RPS model with the low SFR model, it is found that the RPS is effectively 
suppresses SF at $ r < R_{\rm 200}$ if there is some cold gas left in the galaxies. 
While the minor burst makes galaxies galaxies slightly redder than those in the standard model, 
the models with and without the minor burst produces almost the same results. 

Recently, some authors have studied the morphology-density (or radius) relations by using a hybrid method 
of $N$-body simulations and SA models (Okamoto \& Nagashima 2001; Diaferio et al. 2001; Springel et al. 2001). 
In their work, S0 or intermediate $B/T$ population in clusters is hardly reproduced unless a wider $B/T$ range 
is adopted to classify simulated galaxies as S0s. Here we examine how the RPS and minor burst affect the 
intermediate $B/T$ population in clusters. 

We show morphological fractions in the cluster as a function of radius in the righ panels of Fig. \ref{fig2}. 
The simulated galaxies are classified according to their $B$-band $B/T$; $B/T > 0.7$ as bulge-dominated, 
$B/T < 0.4$ as disk-dominated, and $0.4 < B/T < 0.7$ as intermediate. 
By comparison of the CNOC sample with the standard model, it is confirmed that the intermediate fraction
in the standard model is much smaller than the observation, while the bulge-dominated fraction is 
reproduced. Even when we consider the RPS, the intermediate fraction is still too small. 
It is because most of the disk-dominate galaxies are almost pure disk galaxies without minor 
bursts (Okamoto \& Nagashima 2003), 
and then the RPS simply darkens the disk-dominated galaxies rather than increasing their $B/T$s. 
The strange behaviour seen in the RPS models at $r/R_{200} > 0.7$  results from the small number of 
galaxies at the radius because of the spherical high-resolution region. 

The minor bursts increases the intermediate fraction without changing the bulge-dominated fraction 
because it affects only galaxies with small $B/T$s. As a result, the minor burst model reproduces 
the observed fractions quite well despite the fact that it does not influence other properties 
of galaxies. Thanks to minor bursts, now the disk-dominated galaxies tend to be non-pure disk 
galaxies. Consequently, the RPS can change their $B/T$ by the fading of their disks. 
In the minor burst + RPS model, we can see the increasing intermediate fraction toward the centre. 

\section{Summary}

We have investigated the effects of major mergers, 
strangulation, RPS of cold disk gas, and minor mergers on the evolution of
bright cluster galaxies. 
We have used a combination of the cosmological $N$-body
simulations and the SA galaxy formation model.  
This method enables us to study above environmental effects in a fully
cosmological context.  We have determined the model parameters of the
reference model to reproduce galaxy properties at $z = 0$. 

Our results are summarised as follows.
\begin{enumerate}
\item The process that terminates SF in early-type galaxies ($B/T > 0.4$) 
in all environments is starburst trigger by a major merger.  
\item If we adopt appropriate SF time-scale, so as to reproduce the observed 
cold gas mass fraction in the field, the dominant process that determines 
colours of galaxies in the cluster core is the strangulation. 
\item Since the strangulation sufficiently suppresses SF in cluster cores, 
the effect of the RPS is hardly observed. 
\item Minor burst does not affect galaxy properties except for morphology.
\item Without minor bursts the fraction of intermediate $B/T$ galaxies in 
clusters becomes too small.  
The model with the minor burst can reproduce the observed morphology-radius relations.
We conclude that the minor burst is essential process to form intermediate $B/T$ 
galaxies. 
\item The RPS hardly increases intermediate population in the cluster without minor bursts.  
When the minor burst is taken into account, the RPS increases the intermediate 
galaxy fraction in the cluster cores. 
\end{enumerate}

\end{document}